\documentstyle[pmart,twoside,fleqn,epsf]{article} 

\begin{document}

\title{Zero temperature phase diagram
of a $d$-wave superconductor with Anderson impurities}
\author{L. S. Borkowski}
\address{Quantum Physics Division, Faculty of Physics,\\
A. Mickiewicz University,
Umultowska 85, Pozna\'n, Poland}
\date{June 10, 2008}

\maketitle
%
%

\begin{abstract}
We study the model of a $d$-wave superconductor interacting
with finite concentration of Anderson impurities at zero temperature.
The interaction between impurity and conduction electrons
is taken into account within the large-$N$ approximation.
We discuss the obtained phase diagram and its dependence
on the main energy scales.
\end{abstract}

Magnetic and nonmagnetic impurities in correlated electron systems
are an important probe of the properties of the host material. 
There is large and growing body of research on impurities
in high-temperature superconductors.\cite{pjh08,balatsky06}
The refinement of experimental techniques probing local properties
stimulated the theoretical work on defects in those systems.
The investigation of the effects of Zn, Ni and other dopants
in YBCO and BSCCO answered some important questions and raised new ones.

Measurements on some compounds show that the superconducting
order parameter is not uniform over
the entire sample.\cite{cren2000}-\cite{davis2005}
Scanning tunnelling spectroscopy measurement showed that modulation
of the structure of $\rm Bi_2Sr_2CaCu_2O_{8+x}$ is correlated locally with
the magnitude of the energy gap.\cite{slezak07}
The spatial variation of $\Delta_0(\textbf{r})$
may result e.g. from the structural supermodulation affecting
the strength of local pairing interaction.\cite{hirschfeld2008}

The intrinsic spatial variation of the superconducting gap
raises the possibility
of observing impurity states on both
sides of the quantum phase transition in the same sample.
Theoretical work on magnetic impurities in systems
with reduced density of states near the Fermi
surface\cite{wf90}-\cite{yu2002}
showed that the resonant impurity states may be viewed as a sensitive probe
of the superconducting state.
If the coupling to the magnetic impurity is small compared
to the energy scale associated with the gap, the
impurity is decoupled from conduction electrons.
This impurity quantum phase transition
occurs at finite coupling, provided the particle-hole
symmetry is broken, and may be studied by STM techniques.
The low-energy behavior of the model depends
on the exponent
$r$ in the conduction electron density of states,
$N(\epsilon) \sim |\epsilon|^r$, where the Fermi level
is fixed at $\epsilon = 0$.

The interaction of impurities with the conduction
electron band may be studied in the Anderson model
with a BCS-type pairing interaction,

\begin{eqnarray}
\nonumber
H & = & \sum_{k,m} \epsilon_k c^\dagger_{km} c_{km}
+E_0 \sum_m f^\dagger_m f_m
+ V \sum_{k,m} [ c^\dagger_{k,m} f_m b + h.c.]\\
& + &\sum_{k,m} [\Delta(k) c^\dagger_{km} c^\dagger_{-k-m} + h.c.]
\end{eqnarray}

This model allows studying also the mixed valence regime where
the impurity occupation number is less than 1 and charge
fluctuations are dominant.
We assume a two-dimensional $d$-wave order parameter of the form
$\Delta(k) = \Delta_0 \cos(2\phi)$, where $\phi$ is the angle
in the $k_x-k_y$ plane.
The constraint of single occupancy of the impurity site by adding a term
$\lambda (n_f-1)$ to the Hamiltonian, where $\lambda$ is the Lagrange
multiplier and taking $\lambda \rightarrow \infty$.
Minimizing the free energy with respect to the resonant level
energy $\epsilon_f$ and $z=<b^\dagger>=<b>$
and taking the mean field approximation we obtain

\begin{equation}
\label{MF1}
\frac{1}{N} = - \textrm{Im}\int_{-\infty}^\infty
d\omega f(\omega) \textbf{G}_f (\omega+i0^+) ,
\end{equation}

\begin{equation}
\label{MF2}
{\frac{E_0-\epsilon_f}{V^2}}
= \textrm{Im} \int_{-\infty}^\infty d\omega f(\omega)
\textbf{G}^0 (\omega + i0^+)
\textbf{G}_f (\omega + i0^+)] .
\end{equation}

Equations equations (\ref{MF1}) and (\ref{MF2}) are solved
self-consistently with the gap equation

\begin{equation}
\Delta(k) = \int_{-\infty}^{\infty} d\omega
f(\omega) \sum_{k^\prime} V_{kk^\prime}
\textbf{G}(k^\prime,\omega).
\end{equation}

The self-energy in the full conduction electron Green's function $\textbf{G}$
is averaged over impurity positions.
We solve these equations self-consistently and obtain the phase diagram
as a function of the parameters of the model.

\begin{figure}[!]
  \epsfysize=10cm
  \centerline{\epsfbox{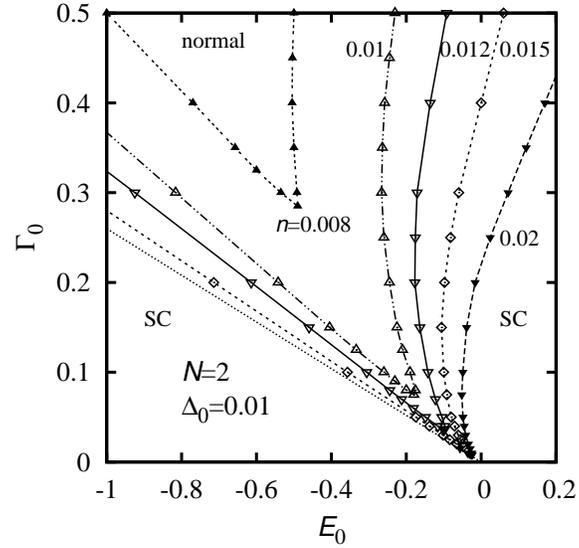}}
\vskip -3.cm
  \caption{The phase diagram of the $d$-wave superconductor with nondegerate Anderson
  impurities for several impurity concentrations. The lines are guide to the eye.
  The order parameter amplitude is $\Delta_0 = 0.01D$, where $D$ is half
  of the conduction electron bandwidth. All energies are scaled
  in units of $D$.
  The dotted line indicates the impurity quantum phase
  transition.}
  \label{fig1}
\end{figure}

The obtained phase diagram is shown in Fig. 1.
The approximate slope of the impurity quantum phase transition line, $-0.26$,
agrees with results of the numerical renormalization group (NRG)
method\cite{ingersent98} and large-$N$
one-impurity calculation.\cite{yu2002}

\begin{figure}[!]
  \epsfysize=11cm
\centerline{\epsfbox{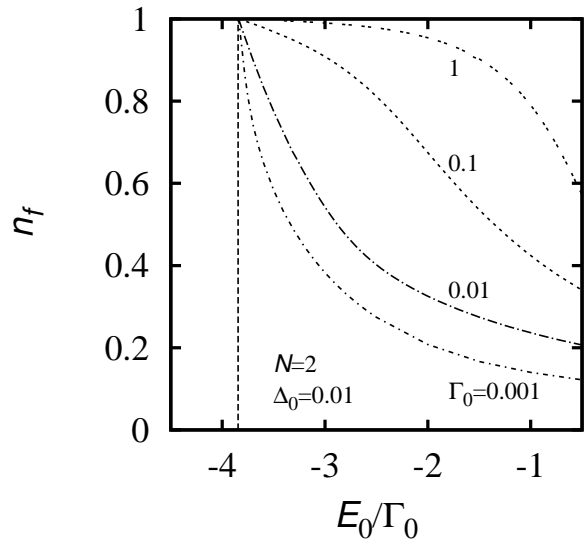}}
\vskip -3.3cm
\caption{The impurity occupation number as a function of the ratio
$E_0/\Gamma_0$ for several values of $\Gamma_0$. 
At the impurity transition $n_f \rightarrow 1$.
}
\label{fig2}
\end{figure}

\begin{figure}[!]
  \epsfysize=10cm
\centerline{\epsfbox{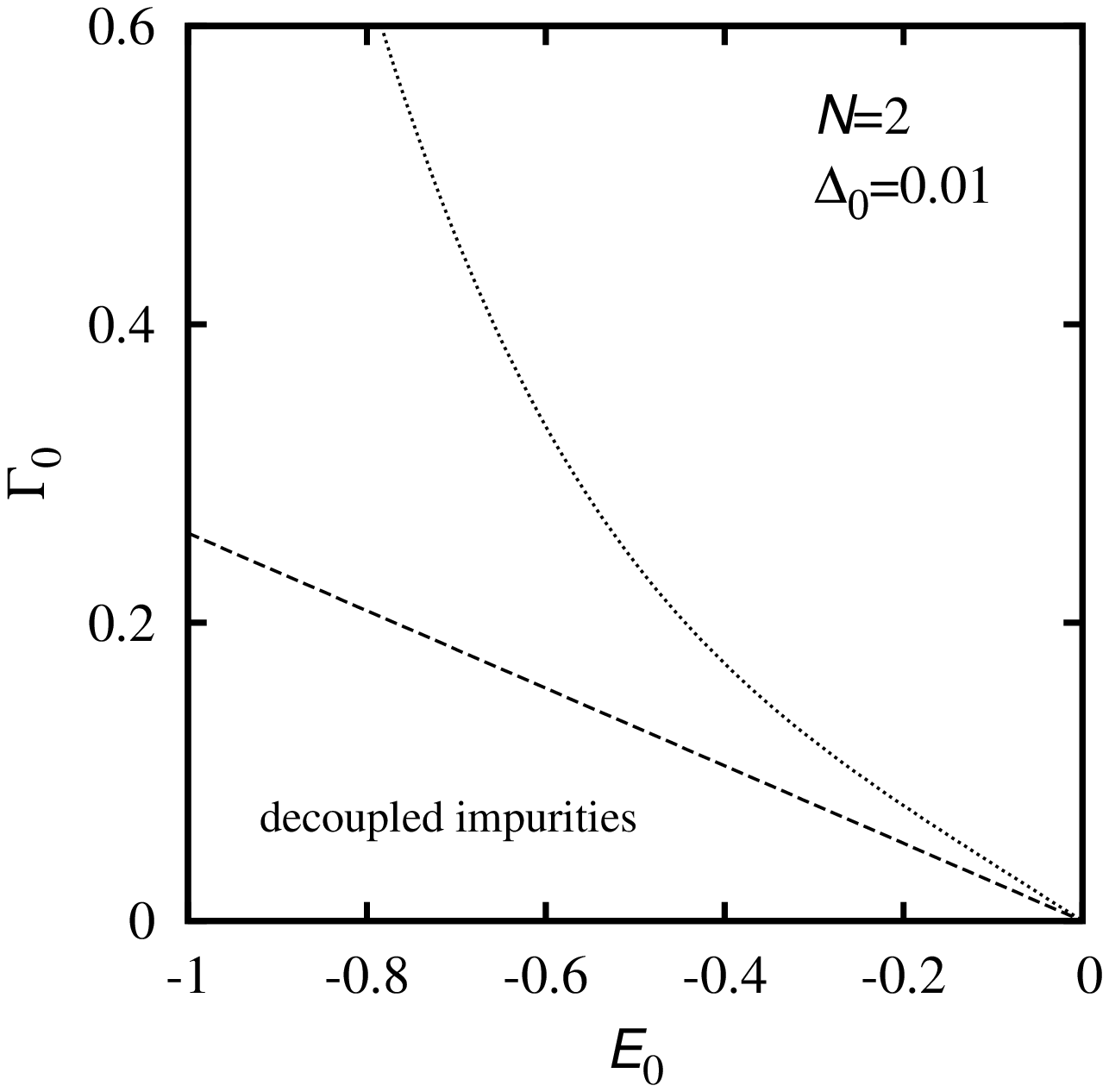}}
\vskip -3cm
\caption{The location of the critical point (dotted line)
and the impurity decoupling transition (broken line).
}
\label{fig3}
\end{figure}

When the bare impurity level $E_0$ lies closer to the Fermi energy
the self-consistent treatment of finite concentration
of impurities leads to reentrant behavior.\cite{lsb2008}
In that part of the phase diagram
the pair-breaking is weaker and is spread over wide energy range.

Fig. 2 shows the impurity occupation number $n_f$ at the impurity
decoupling transition. It reaches $1$ at the transition.
For $\Gamma_0 \gg \Delta_0$, $n_f(E_0)$ slowly
approaches $1$, $|dn_f/dE_0|_{\Gamma_0=\textrm{const}} \ll 1$,
as $E_0/\Gamma_0 \rightarrow
(E_0/\Gamma_0)_{\textrm {critical}}$.
However for small $\Gamma_0$ the dependence of $n_f$ on
$E_0/\Gamma_0$ becomes singular and the slope
$|dn_f/dE_0|_{\Gamma_0=\textrm{const}} \gg 1$.

In the limit of vanishing interaction $\Gamma_0 \rightarrow 0$,
$E_0=0$ is the singular point of the model.
For $E_0 > 0$ the superconducting state survives
even for large impurity concentration.

For any finite $n$ there is a critical point $(\Gamma_{0c}, E_{0c})$.
For $\Gamma_0$ slightly larger than $\Gamma_{0c}$
there is a superconductor-normal
state transition at $E_{01} < E_{0c}$ and another normal-superconductor
transition at $E_{02}$, where $E_{01} < E_{02} < E_{0c}$.
At fixed $n$, $E_{02}-E_{01}
\simeq \alpha(n) (\Gamma_0-\Gamma_{0c})$, where $\alpha(n)$
weakly depends on $n$.
The location of this critical point in the $E_0$-$\Gamma_0$ plane
is shown in Fig. 3.

It would be interesting to test this theoretical picture
in experiment. Near the impurity quantum phase transition
the impurity state is very sensitive
to small changes of hybridization $\Gamma_0$ or impurity level
energy $E_0$. In compounds with spatially varying energy gap
this could lead to impurities existing on the two sides
of the transition line in different parts of the sample.

The phase diagram calculated in this work
might also be relevant in some heavy-fermion compounds where
similar competition occurs between energy scales associated with the Kondo
screening and the superconducting correlations.
Studies of $\rm CeCu_2(Si_{1-x}Ge_x)_2$
under varying hydrostatic pressure reveal two superconducting
domes in the phase diagram.\cite{steglich2004,steglich2008}
The existing interpretation of this dependence on pressure
relies on additional valence-fluctuation
mediated pairing mechanism.\cite{miyake2002}

However our work suggests that the second superconducting dome
in \\ $\rm CeCu_2(Si_{1-x}Ge_x)_2$
at high pressure may follow from
weakened pair-breaking.
The change of pressure shifts the chemical potential and brings
the system to the mixed-valence regime when
the bare $f$-level $E_0$ of Ce ions approaches $E_F$.
The phase diagram in Fig. 1 shows that in this limit
superconducting correlations are less affected by pair-breaking.

The large-$N$ method used in the present calculation
gives qualitatively similar results for larger $N$.
Also the symmetry of the order parameter should not
introduce drastic changes to the phase diagram. The reentrant
behavior as a function of $E_0$ results from the competition
between the formation of the impurity resonance and
superconducting correlations and depends mainly
on the ratio of the relevant energy scales.

Extension of the theory beyond the mean field
should not change the phase diagram qualitatively.
A more detailed description of physics
in the vicinity of the impurity transition line
requires careful treatment
of low-energy scattering in specific superconducting
compounds.

\end{document}